\documentclass[preprint,12pt]{elsarticle}

\usepackage{svg}
\usepackage{graphicx} 
\usepackage{amsmath}
\usepackage{amssymb}

\usepackage{xcolor}
\usepackage{float}
\usepackage{svg}

\usepackage{graphicx}% Include figure files
\usepackage{dcolumn}% Align table columns on decimal point
\usepackage{bm}% bold math
\usepackage[utf8]{inputenc}
\usepackage[T1]{fontenc}
\usepackage{mathptmx}
\usepackage{etoolbox}

\journal{}

\begin{document}
	\begin{frontmatter}
		\title{Evolving Neural Networks Reveal Emergent Collective Behavior from Minimal Agent Interactions}
		\author{G. S. Y. Giardini$^{a}$}
        \author{J. F. Hardy II$^{a}$}
		\author{C. R. da Cunha$^{a}$}
		\ead{carlo.cunha@nau.edu}
		\affiliation{organization= {School of Informatics, Computing, and Cyber-Systems, Northern Arizona University}, 
			addressline={1295 S. Knoles Dr.},
			city={Flagstaff},
			state={AZ},
			postcode={86011},
			country={USA}			
		}
		 
		\date{\today}
		
		\begin{abstract}
			Understanding the mechanisms behind emergent behaviors in multi-agent systems is critical for advancing fields such as swarm robotics and artificial intelligence. In this study, we investigate how neural networks evolve to control agents' behavior in a dynamic environment, focusing on the relationship between the network's complexity and collective behavior patterns. By performing quantitative and qualitative analyses, we demonstrate that the degree of network non-linearity correlates with the complexity of emergent behaviors. Simpler behaviors, such as lane formation and laminar flow, are characterized by more linear network operations, while complex behaviors like swarming and flocking show highly non-linear neural processing. Moreover, specific environmental parameters, such as moderate noise, broader field of view, and lower agent density, promote the evolution of non-linear networks that drive richer, more intricate collective behaviors. These results highlight the importance of tuning evolutionary conditions to induce desired behaviors in multi-agent systems, offering new pathways for optimizing coordination in autonomous swarms. Our findings contribute to a deeper understanding of how neural mechanisms influence collective dynamics, with implications for the design of intelligent, self-organizing systems.
		\end{abstract}
		

\end{frontmatter}

\section{Introduction}
Many physical systems display properties that are absent in their components. From the atomic level \cite{Anderson1972} to the biological \cite{Alberts1994} and social realms \cite{Couzin2007}, we see countless examples of systems exhibiting emergent properties \cite{Gershenson2020,MunozJose2022}.
This concept also applies to artificial environments like cities where the interactions of countless individuals result in a dynamic structure that grows, consumes energy, and continually reshapes itself \cite{Bettencourt2007}. Understanding these natural and social examples of emergence can inform the design of artificial systems with practical applications. For instance, drones used in search and rescue missions, environmental monitoring, or disaster management must coordinate effectively while avoiding collisions \cite{FLOREANO2023}. 
By applying our understanding of emergent behavior, we can design agents that self-organize and adapt in dynamic environments, forming a collective intelligence \cite{Suran2020}.

Several approaches model emergent behaviors in multi-agent systems. Early attempts, for example, introduce a polarized motion to point-like agents that interact with their neighbors according to fixed rules \cite{Bajec2007}. The consecrated Vicsek model \cite{Vicsek1995}, on the other hand, uses the alignment with nearest neighbors as a fundamental rule. Other attempts rely on game theoretical strategies as an approach for collective emergence \cite{Pacheco2009,Perc2013}. 

Finding the minimal and primitive set of interaction rules capable of producing emergent behavior, however, is a challenging task. In most existing models, the interaction is given \textit{a priori} without explaining if these rules form a minimal or primitive set \cite{Bajec2007}.
This limitation is sometimes solved via reinforcement learning techniques \cite{Bellman1954} where agents learn optimal behaviors by adjusting strategies depending on rewards and penalties provided by their environment. This process can lead to agents that eventually develop coordinated collective behavior. However, reinforcement learning usually requires precise knowledge about the environment. Moreover, the search space in natural systems can be huge, which limits exploration \cite{Padakandla2020}.

Evolutionary programming is a competing strategy where collective dynamics arise naturally. Under evolutionary pressure, agents pass their most adapted features to offspring through reproduction and mutation \cite{Kwasnicka2011,Ramos2019}. These and similar approaches are often complex, making heuristic analysis complicated \cite{Charlesworth2019}.

Our objective in this work is to investigate minimal and primitive conditions that lead to collective and spontaneous self-organization in decentralized systems. For this, we draw inspiration from natural processes \cite{Holland1998,Dorigo1999,Lecun2015} and propose a model of minimally intelligent agents endowed with explainable shallow neural networks used to choose a movement direction. These networks are trained in an evolutionary algorithm with one of the simplest evolutionary pressures: proximity. This mimics the behavior of social animals and serves as a foundation for other, more complex behaviors such as protection \cite{Hamilton1971,Schaik1983}, reproduction \cite{Schaik1983}, hunting \cite{Packer1988}, and navigation \cite{Couzin2005,Majolo2022}. We also introduce other simple physical constraints like volume exclusion, rotational inertia, and a field of vision that limits the angle at which agents can detect their neighbors. These three assumptions ensure physical realism, as all animals have limited fields of view, cannot rotate impossibly fast, nor can occupy the same space already occupied by another. By limiting the rules to these foundational elements, we seek to explore if proximity alone can serve as a precursor to collective behavior purely through local interactions and evolutionary dynamics.  

Having minimally intelligent agents allows us to determine qualitatively, how each parameter of their neural networks influences the final dynamics of the system and how the agents interact with their neighbors \cite{Barron1993,Bengio2007}.

Next, we detail our model, outline the specifics of the neural networks, and explain the interaction mechanism between agents. We then present metrics to verify the emerging collective behavior. Finally, we analyze our findings, verifying key points in the emergence of collective motion, and then discuss possible reasons that lead to emergent behavior.

\section{Computational Methods}
    \subsection{Mathematical Model}
    Our model is formalized by the tuple $\nu=(N,\mathbf{X},\mathcal{L},L,\mathbf{N},\mathcal{N},\delta,R_{col},\phi,\omega,\varphi)$, where $N$ is the number of agents in the system, $\mathbf{X}=\{\mathbf{x}_1,\mathbf{x}_2,\hdots,\mathbf{x}_N\}$ is the set of positions of the agents. Each position $\mathbf{x}_i(t)\in\mathcal{L}$, where $\mathcal{L}\subset\mathbb{Z}^2$ is a finite $L\times L$ lattice. $\mathbf{N}=\{n_1,n_2,\hdots,n_N\}$ is a set of shallow neural networks \cite{Rosenblatt1958,Cybenko1989,daCunha2024} whose outputs give the directions the polarized agents move. These neural networks mimic some level of intelligence found in most complex living organisms \cite{Wasserman2009}. Even some cells, the biological units of life, have been shown behaviors analogous to intelligence \cite{Tang2018}.
$\mathcal{N}=\{\mathbf{y}_1,\mathbf{y}_2,\hdots,\mathbf{y}_6\}$ is a ranked metric neighborhood consisting of the $6$ nearest neighbors. For a metric function $d$, and positions $\mathbf{y}_{1<i\leq 6}$ around $\mathbf{x}$, $d(\mathbf{x},\mathbf{y}_1)<d(\mathbf{x},\mathbf{y}_2)<\hdots<d(\mathbf{x},\mathbf{y}_6)$. This is based on the fact that many animals interact with a fixed number of neighbors rather than with all neighbors within an interaction radius \cite{Faria2010,Read2011,Bialek2012,Gautrais2012}, for birds the number of nearest neighbors was verified to be $6 \sim 7$ \cite{Ballerini2008}. 

\begin{figure}
    \centering
    \includegraphics[scale=0.2]{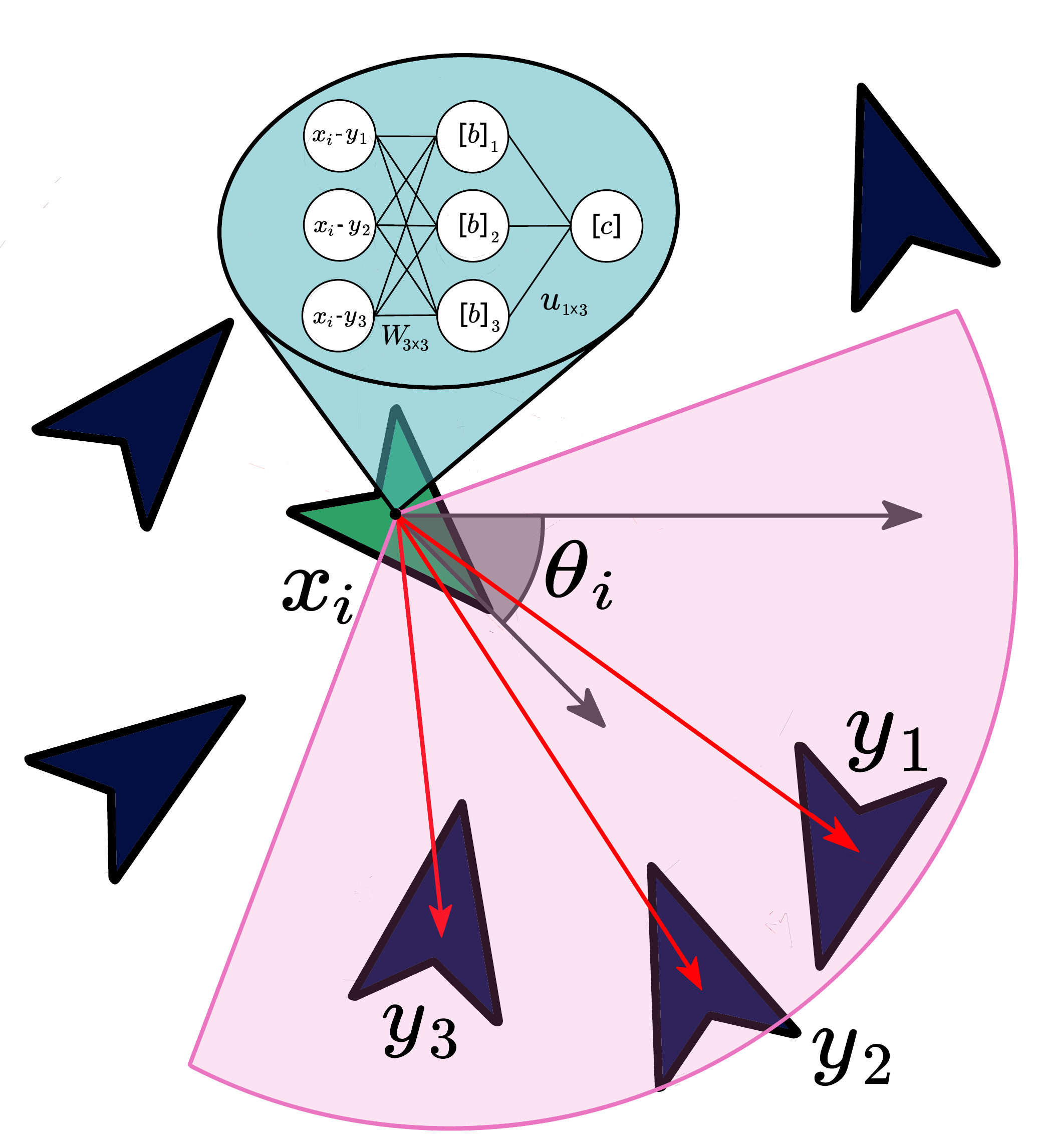}
    \caption{Simplified diagrammatic representation of interacting agents with 3 nearest neighbors. In the diagram, the number of nearest neighbors has been defined as $k=3$ for simplicity. The variables $\mathbf{x}_{i}$ and $\theta_{i}$ are respectively agent $i$'s position and orientation, the pink sector of a circle represents its field of view. The variables $\mathbf{y}_{j}$ are the positions of neighboring agents inside agent $i$'s field of view, and the blue region denotes its neural network, with inputs $\mathbf{v} = [\mathbf{x}-\mathbf{y}_{i}]^{3}_{i=1}$, a weight matrix $\mathbf{W}$ with bias vector $\mathbf{b}$ whose result passes through an activation function $tanh$ and is fed into an output layer with a weight vector $\mathbf{u}$ and activation value $c$.}
    \label{fig:sec2}
\end{figure}

The input of the neural network for an agent located at $\mathbf{x}$ is a vector of differences $\mathbf{v}=[\mathbf{x}-\mathbf{y}_i]^6_{i=1}$, where $\mathbf{y}_i$ are the elements in the ranked neighborhood of $\mathbf{x}$. Also, only agents inside one's field of vision $\phi\in[0,2\pi]$ are considered \cite{Costanzo2018}. Our model is illustrated in Fig. \ref{fig:sec2}.

The neural network is composed of a single hidden layer with $8$ neurons given by:

\begin{equation}
    \mathbf{h}=\text{tanh}\left(\mathbf{W}\mathbf{x}+\mathbf{b}\right),
\end{equation}
where $\mathbf{W}$ is a weight matrix and $\mathbf{b}$ is a bias vector. This feeds into an output layer described by:

\begin{equation}
    Y=\text{tanh}\left(\mathbf{u}\cdot\mathbf{h}+c\right),
\end{equation}
where $\mathbf{u}$ is a weight vector and $c$ is a bias value.

In our model, $\delta:\mathbf{X}\times\mathbf{X}$ is a transition function that adjusts the agent's angle based on the output of its neural network:

\begin{equation}
    \theta(t+\Delta_t) = \begin{cases}(1-\lambda)Y\pi+\lambda\xi_t & \text{, if } |\theta(t+1)-\theta(t)|<\omega\Delta_t\\
    \text{sgn}\left(\theta(t+1)-\theta(t)\right)\omega\Delta_t & \text{, otherwise.}\end{cases}
\end{equation}
In this equation, $\Delta_t$ is the time step, $\lambda\in[0,1]$ adjusts the intensity of a noise level applied to this angle, and $\xi_t$ is a Wiener process \cite{Wiener1923,Genthon2020,daCunha2022}. The noise level emulates random factors that are external to the agent like variations in the weather. Additionally, $\omega$ is the maximum angular speed \cite{Costanzo2018}.

The agent translates this new angle into a new position according to:

\begin{equation}
    \mathbf{r}(t+\Delta_t)=\begin{cases}\mathbf{r}(t)+\begin{bmatrix}\cos(\theta)\\ \sin(\theta)\end{bmatrix}v_0\Delta_t & \text{, if }|\mathbf{r}(t+1)-\mathbf{s}(r+1)|<R_{col}\forall {\mathbf{s}\neq\mathbf{r}}\\
    \mathbf{r}(t) & \text{, otherwise}\end{cases}.
\end{equation}
In this equation, $v_0$ is the velocity amplitude and $R_{col}$ is a collision radius.

The parameters $(\mathbf{W},\mathbf{u},b,c)$ of each neural network are trained online using an evolutionary approach. After every period $T=300$ steps, two agents are selected based on fitness-proportionate selection \cite{Lipowski2012}, with agents possessing higher fitness values being more likely to be chosen. These selected agent's parameters, that are flattened into vectors within the parameter space, are mixed to create a new neural network according to uniform crossover \cite{daCunha2024}. The parameter space is the set of all possible vectors obtained by flattening the neural network parameters. The newly created network is subject to mutation and is then assigned to the agent with the lowest fitness score, ensuring an elitist selection mechanism \cite{Holland1998,daCunha2024}. This process guarantees that only the least-fit agents are replaced by offspring derived from the more successful individuals, promoting convergence towards optimal solutions \cite{Holland1998,Yao1999,FLOREANO2023,daCunha2024}. The procedure is repeated until 20\% of the population has reproduced, at which point the period of $T=300$ steps is restarted \cite{FLOREANO2023}.

In our model, we define the fitness score as the average proximity to an agent's $k$-nearest neighbors. This fitness score represents the tendency to stay close to neighbors. Mathematically, the fitness is computed as the  average of the distances to the $k$-nearest neighbors:

\begin{eqnarray}
    \varphi_{i}(t+\Delta_t) = \frac{1}{t+\Delta_t} \left[t \, \varphi_{i}(t) + \frac{\Delta_t}{k} \sum_{j\in \mathcal{N}_{i}} |\mathbf{x}_i-\mathbf{x}_j|^{-1} \right].
    \label{eq:Sec_2_fitness}
\end{eqnarray}

The system can eventually get stuck in a local minimum of the fitness landscape. In this state, all agents can possess similar neural networks that are insensitive to input changes. This lack of variability is analogous to a population facing extinction due to the absence of genetic diversity. To prevent this situation, we introduce fitness sharing, a method that punishes individuals who are too similar \cite{FLOREANO2023}. 
Mathematically, the shared fitness is defined as: 

\begin{eqnarray}
    \varphi_{i}^{sh}(t) = \frac{\varphi_{i}(t)}{\sum_{j=1}^{N}sh(\Delta_{ij})},
\end{eqnarray}
where:

\begin{eqnarray}
    sh(\Delta_{ij}) = 
        \begin{cases}
        1 - \left(\frac{\Delta_{ij}}{\sigma_{\text{share}}}\right)^{\alpha}, & \text{if } \Delta_{ij}<\sigma_{\text{share}}\\
        0, & \text{if } \Delta_{ij}\geq\sigma_{\text{share}}
        \end{cases}.
\end{eqnarray}

The latter is the sharing function, a similarity measure between neural networks. It depends on distances $\Delta_{ij}$ in the neural network's parameter space. We used $\alpha=1$ and $\sigma_{\text{share}} = \beta\, \langle\Delta_{ij}\rangle_{ij}$. We set $\beta=0.38$, as it provides a balance between training speed and avoiding local minima. 

The training process is carried out until it reaches a steady state around $2\, 10^{6}$ steps. The agents' neural networks are randomly initialized.

    \subsection{Data Extraction}
    \label{sec:Measured_Quantities}
We employed a set of computational methods to analyze the behavior of autonomous agents and the emergent phenomena observed in the swarm. These methods capture global and local dynamics, including alignment, clustering, and collective motion. Below, we describe the primary observables and metrics that quantify these behaviors.

First, we track fitness over time ($\varphi(t)$) \cite{FLOREANO2023}, which reflects the evolutionary training progress and gives insight into the agents' proximity. To complement this, Vicsek's order parameter ($\Psi$) \cite{Vicsek1995} was used to determine the degree of alignment among agents. This metric captures the extent to which agents move in the same direction. 
Additionally, we draw an analogy to fluid dynamics by introducing relative viscosity to quantify how orientation changes affect the system’s cohesion. We compute relative viscosity using a modified version of Ubbelohde's formula \cite{Ubbelohde1937}, comparing the time agents take to traverse a given distance ($t$) with that of a reference fluid in laminar flow ($t_0$):

\begin{eqnarray}
    \eta_{\text{r}} = \frac{t\, \rho}{t_{0}\, \rho_{0}},
\end{eqnarray}
High relative viscosity values suggest a disorderly system with frequent reorientation, while low values indicate stable, cohesive movement.

Additionally, we perform multi-variable linear regression to explore the relationship between neural network inputs and outputs. By assessing the regression quality using the coefficient of determination ($R^2$), we evaluate the non-linearity of the trained networks. This analysis uses Julia’s GLM library\footnote{https://juliastats.org/GLM.jl/stable/}\cite{Bezanson2017}. 

To further study migration patterns, we constructed connectivity graphs. In these graphs, each agent in a group is represented by a node in $V\subseteq\{1,\dots,N\}$. The connectivity graph $G=(V,E)$ is formed by creating edges  $E\subseteq\{(v_i,v_j)\in V\mid|\mathbf{x}_i-\mathbf{x}_j|\leq R_g, i,j\in V\}$ between agents if their separation is less than $R_g=2.5$ grid units. The complexity of a migration pattern is measured by the spectral gap (first nonzero eigenvalue) of the Laplacian matrix, computed for the largest connected component of the connectivity graph. Large spectral gaps indicate more connected and simple graphs, while small spectral gaps imply less connected and more complex graphs.

Finally, we construct a second graph representing the neural network of the fittest agent for each stable solution reached by the evolutionary algorithm. This visualization provides a qualitative assessment of how agent decision-making is influenced by its neighbors. The graph’s nodes represent the inputs and outputs of the NN, while the edges depict the strength of connections, visually encoded by color and line thickness. 

\begin{table}[!ht]
    \centering
    \begin{tabular}{c c c}
        \hline
        Parameters & Name & Parameter range \\  
        \hline
        N & Number of agents & 100-200\\
        $L^2$ & Lattice area & 20 $\times$ 20 area units\\
        $\rho = N / L^{2}$ & Agent density & $0.25-0.5$ \\
        $v_{0}$ & Agent velocity & 1 length units $(\text{time units})^{-1}$ \\
        %$\xi_{i}$ & Input noise \\
        $\xi$ & Output noise & $0.0 - 1.0\, rads$ \\
        R & Collision radius & $0.0 - 0.2 \text{ length units}$ \\
        FoV & Agent field of vision & $0-2\pi\, rads$ \\
        $\omega_{\text{max}}$ & Maximum turning angle per time-step & $0-2\pi$ $\Delta t^{-1} \, rads$  \\
        $\kappa$ & Number of nearest neighbors & 6 \\
        \hline
    \end{tabular}
    \caption{Table of parameters for the multi-cognitive agent model.}
    \label{tab:sec3-parameters_table}
\end{table}
 
Table \ref{tab:sec3-parameters_table} summarizes the parameters used in our analyses. In all analyses, we constrained the parameter space by keeping the system size $L$ fixed while varying the number of agents $N$. The agent density $\rho$ was limited to values below $0.5$, as higher densities produced no new behaviors of interest beyond laminar flow, a conclusion reached after analyzing 170 simulations with different density values. Additionally, the velocity of the agents was held constant at $v_{0} = 1.0$ throughout the simulations.

\section{Results}
    \subsection{Training}
\begin{figure}[!ht]
    \centering
    \includegraphics[width=0.9\linewidth]{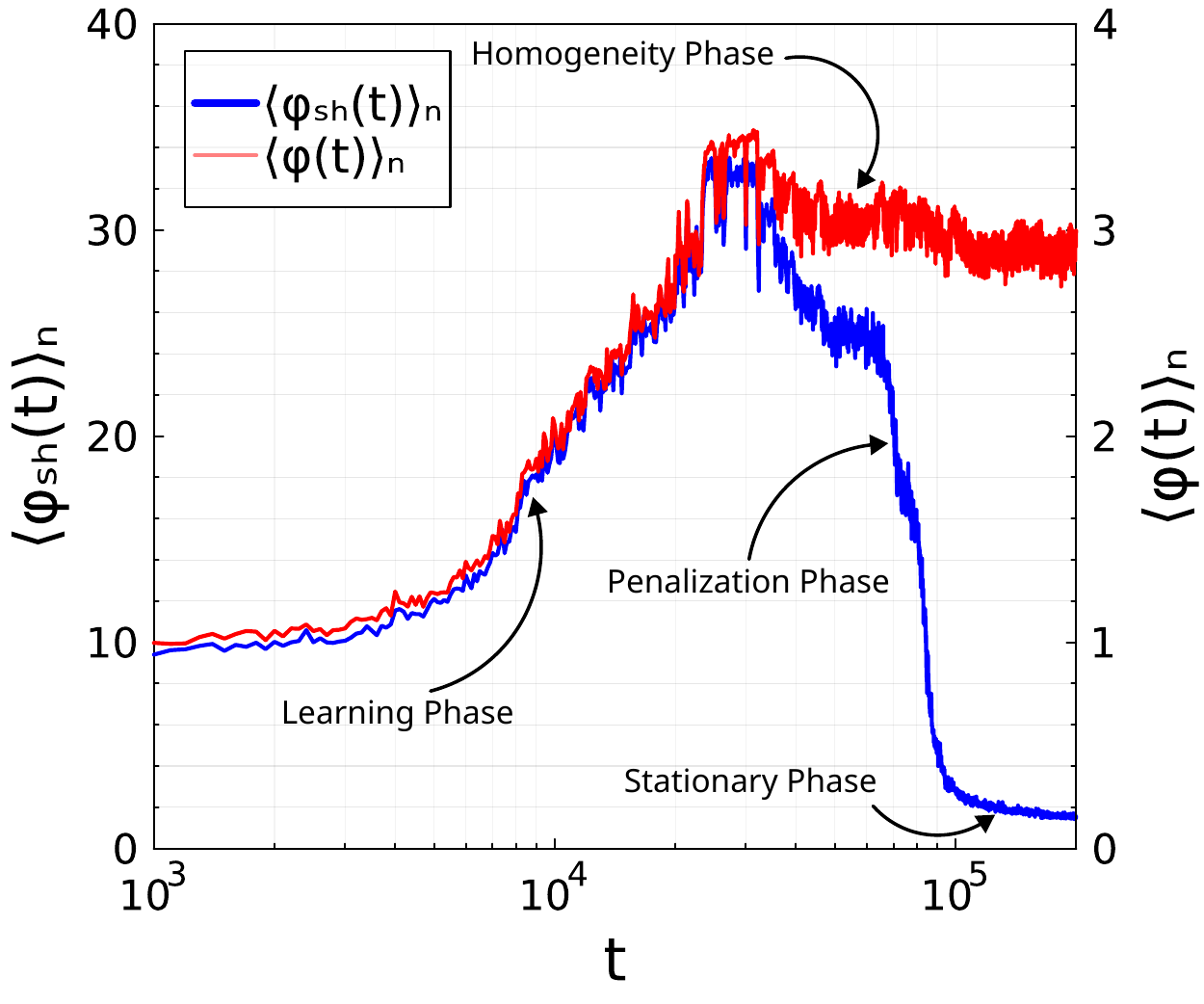}
    \caption{Average fitness ($\langle\varphi(t)\rangle$) and average shared fitness ($\langle\varphi_{sh}(r)\rangle$) as a function of training steps.}
    \label{fig:sec3-Fitness_Evolution}
\end{figure}

    The overall fitness indicates how well the agents are evolutionarily adjusted to the environment, showing a progressive increase, as depicted in Fig. \ref{fig:sec3-Fitness_Evolution}. The shared fitness, however, is defined as the overall fitness normalized by the complement of the average distance between the neural networks in parameter space. If evolution leads to an increase in overall fitness at the cost of declining diversity, the shared fitness decreases. The shared fitness increases at the beginning of the training process but eventually saturates, indicating the homogenization of agents (homogeneity phase). Shared fitness makes more diverse agents more likely to be chosen for reproduction and increases the variability of the population (penalization phase). Eventually, the system increases its diversity of agents to a point that compensates the pressure for homogenization (stationary phase).

    \subsection{Emerging Patterns}
    %======== EFFECTS OF DENSITY
% milling - \cite{Couzin2002}
% effects of topology on milling - \cite{Costanzo2018}
Incorporating collision avoidance was essential for observing the emergence of complex patterns. This adjustment also makes the system more closely resemble natural behaviors, where agents maintain personal space and avoid overlap \cite{Giardina2008, Couzin2002, Couzin2005}. Without this treatment, emergent patterns like flocks and lanes would only form in the presence of external perturbations.

With collision handling introduced, we systematically explored the parameter space described in Tab. \ref{tab:sec3-parameters_table}. Varying the system density, external noise, field of view, and maximum turning angle produced distinct migration patterns. Moreover, we observed order-disorder transitions by increasing noise. 

\begin{figure}
    \centering
    \includegraphics[width=1.0\linewidth]{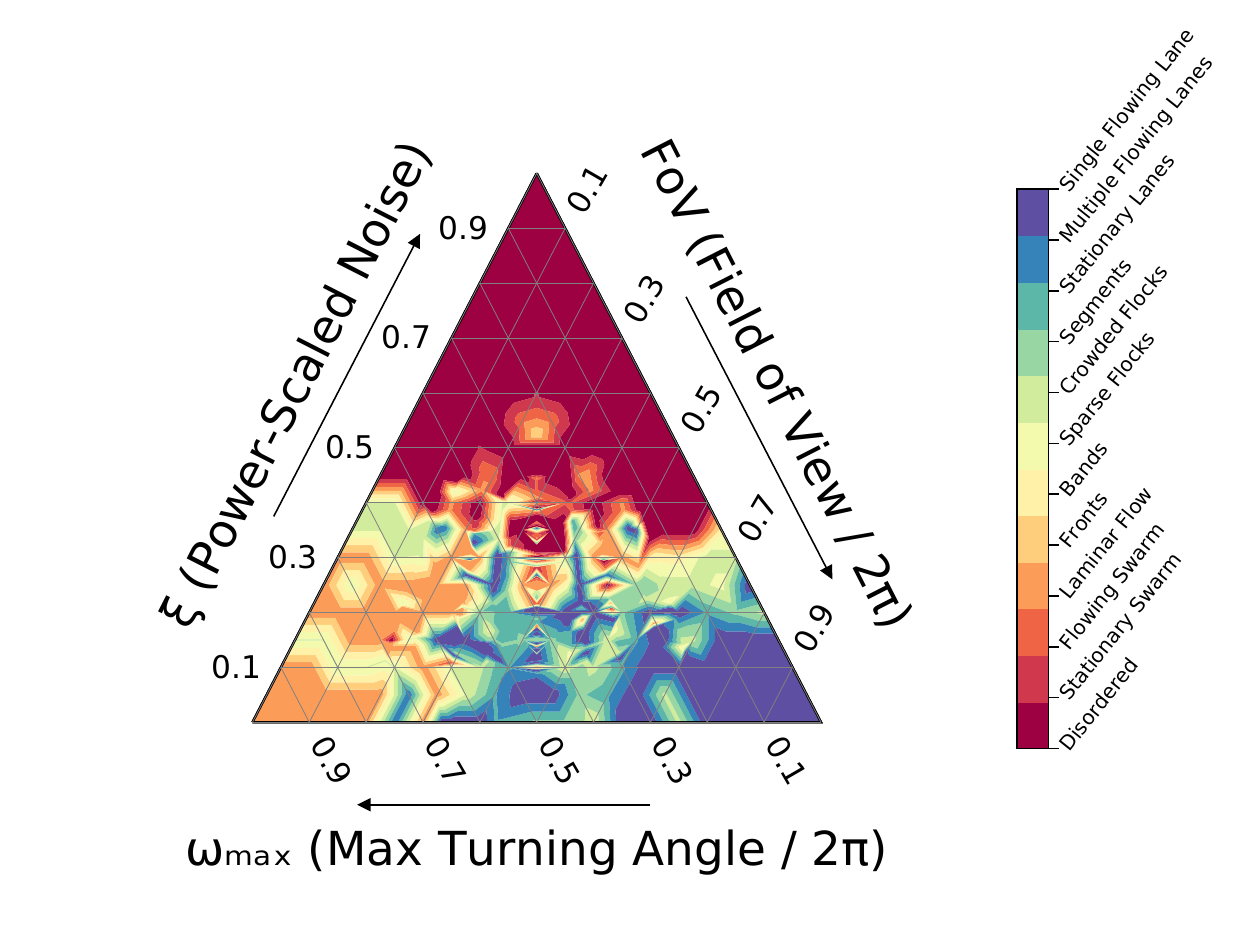}
    \caption{Migration regimes for different configurations of maximum turning angle, rotational noise, and field of vision.}
    \label{fig:sec3_Density_Phase_Diagram}
\end{figure}

We examine the effects of external noise, field of view, and maximum turning angle on the emergence of migration patterns. Results were obtained over a set of at least $336$ different simulations and displayed in ternary phase diagrams. Each point in these graphs corresponds to a specific combination of three independent variables $a_{i\in[1,2,3]}$ that always add up to $1$. A function of these parameters is represented as a color gradient, allowing direct visualization of how the output changes with different combinations of parameters. The results are shown in Fig. \ref{fig:sec3_Density_Phase_Diagram}.

The diagram is divided into distinct regions, each representing a different phase of the system. As the external noise increases, the system transitions between different phases from disordered to structured patterns: i) stationary swarm, characterized by a state where entities form loosely organized clusters and whose group's center of mass remains stationary, ii) flowing swarms share the same definition but with a non-stationary group's center of mass, iii) laminar flow, represented by laminar flows of groups without any movement internal to the group, iv) fronts and bands, described by propagating waves, iv) sparse flocks, given by groups that are constituted by many small flocks, v) crowded flocks, characterized by large, highly organized groups, vi) segments, describe by structured movement in linear arrangements, vii) stationary lane shown as a structured path whose overall center of mass is stationary, viii) multiple flowing lanes, represented by organized multiple structured paths, and ix) single flowing lane, illustrated by a single structured movement.

\begin{figure}[h!]
    \centering
    \includegraphics[width=1.0\linewidth]{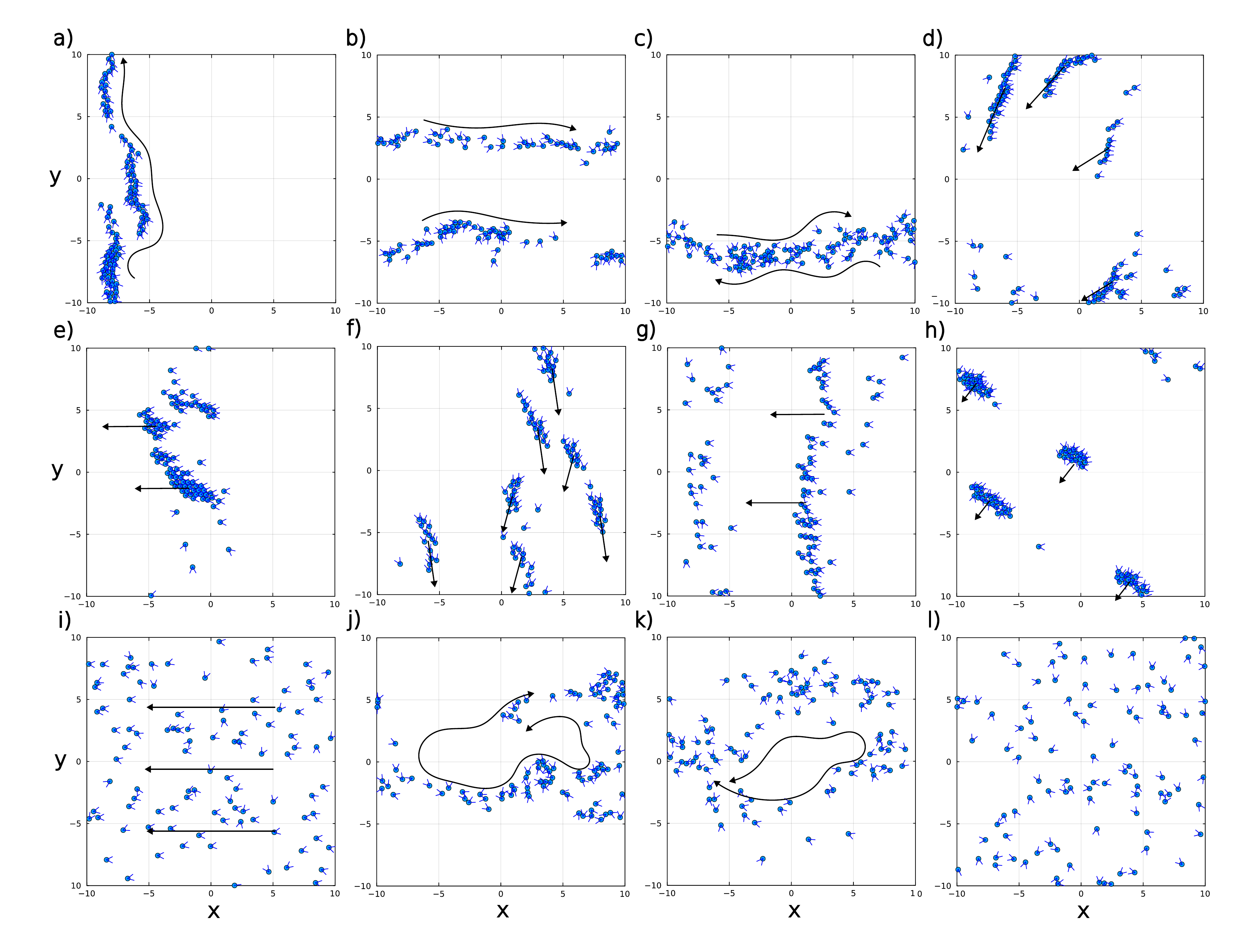}
    \caption{Different patterns obtained by exploring the parameter space: a) single flowing lane, b) multiple flowing lanes, c) stationary lanes, d) segments, e) crowded flocks, f) sparse flocks, g) bands, h) fronts, i) laminar flow, j) flowing swarm, k) stationary swarm, l) disordered. Arrows represent the overall flow direction.}
    \label{fig:sec3_Migration_Types}
\end{figure}

Figure \ref{fig:sec3_Migration_Types} provides visual examples of each migration regime. 

\begin{figure}[h!]
    \centering
    \includegraphics[width=0.9\linewidth]{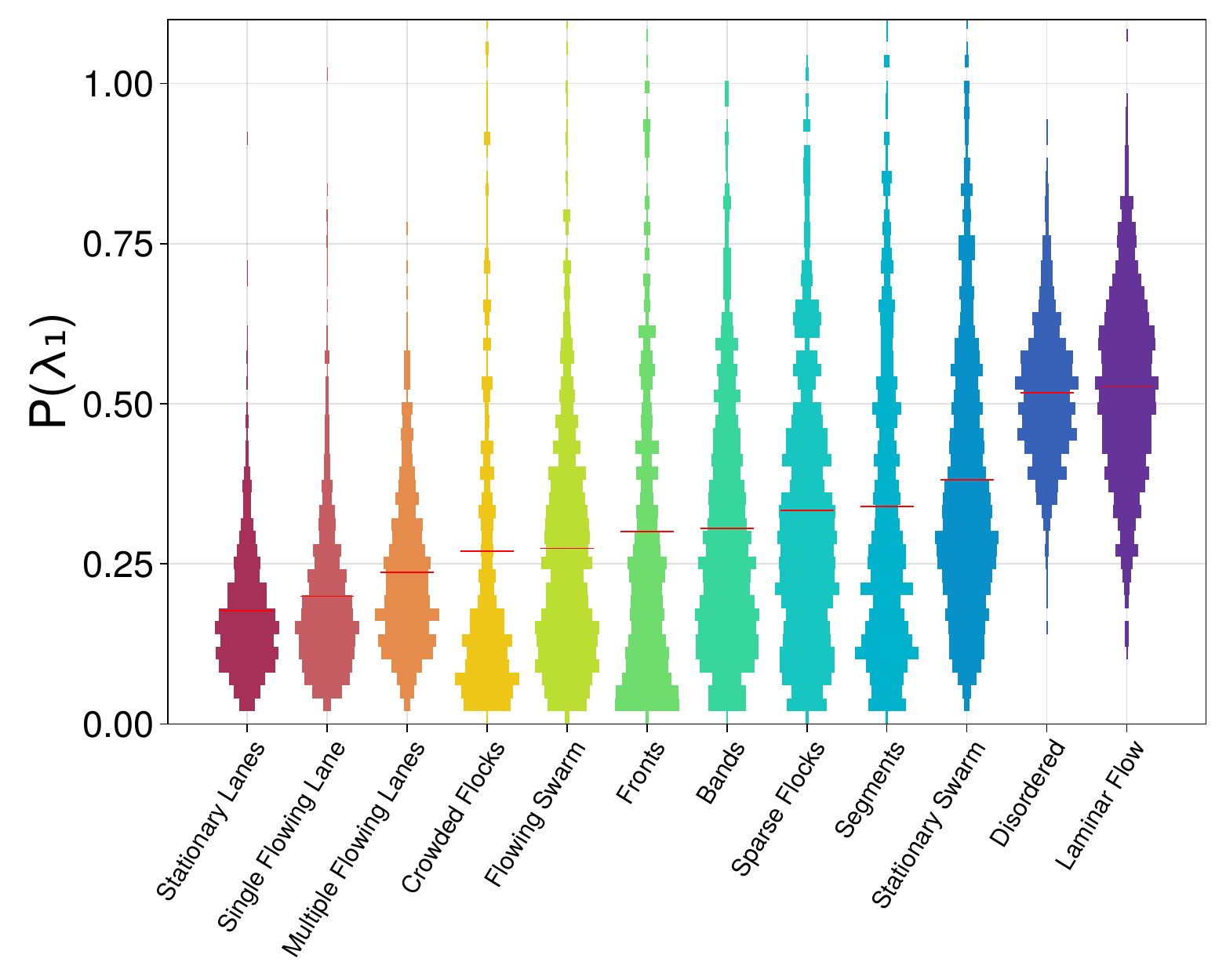}
    \caption{Probability density of the spectral gap time distribution of the connectivity graphs.  Different colors represent distinct emergent patterns, and the red lines indicate the mean of each distribution.}
    \label{fig:sec3_Spectral_Gap_Violin_Plots}
\end{figure}

Table \ref{tab:sec3_Phase_Complexity} presents three measures of complexity for each migration pattern: the first and second moments of the spectral gap time distribution, and the Kullback-Leibler (KL) divergence \cite{Kullback1951} between this distribution and a corresponding Gaussian distribution with the same moments. Notably, the KL divergence is approximately $70$ \% anti-correlated with the first moment but is only weakly anti-correlated ($\sim26$ \%) with the second moment. Therefore, less complex patterns, characterized by large average spectral gaps and high connectivity, tend to be more Gaussian distributed, indicating more temporal regularity. Figure \ref{fig:sec3_Spectral_Gap_Violin_Plots} shows corresponding violin plots to this data.

\begin{table}[h!]
    \centering
    \begin{tabular}{c c c c}
        \hline 
        Emergent pattern & $\mu(\lambda_{1})$ & $\sigma(\lambda_{1})$ & KL Divergence \\
        \hline
        Stationary lane          & 0.1765 & 0.1020 & 7.0341 \\
        Single flowing lane      & 0.1994 & 0.1305 & 3.5308 \\
        Multiple flowing lanes   & 0.2367 & 0.1258 & 5.1305 \\
        Crowded flocks           & 0.2702 & 0.3153 & 4.0780 \\
        Flowing swarm            & 0.2741 & 0.1920 & 1.9723 \\
        Fronts                   & 0.3001 & 0.3051 & 2.9593 \\
        Bands                    & 0.3052 & 0.2066 & 1.8040 \\
        Sparse flocks            & 0.3333 & 0.2145 & 1.4380 \\
        Segments                 & 0.3399 & 0.2571 & 2.5028 \\
        Stationary swarm         & 0.3811 & 0.2136 & 1.4304 \\
        Disordered               & 0.5174 & 0.1141 & 1.7852 \\
        Laminar-flow             & 0.5263 & 0.1527 & 1.4252 \\
        \hline
    \end{tabular}
    \caption{Three measures of complexity for the migration patterns: the first $\mu(\lambda_{1})$ and second moments $\sigma(\lambda_{1})$ of the spectral gap time distribution, and the Kullback-Leibler (KL) divergence between this distribution and a corresponding distribution with the same moments. The distributions were computed over $1000$ measurements.}
    \label{tab:sec3_Phase_Complexity}
\end{table}

    \subsection{Effects of Noise}
    
\begin{figure}
    \centering
    \includegraphics[width=0.8\linewidth]{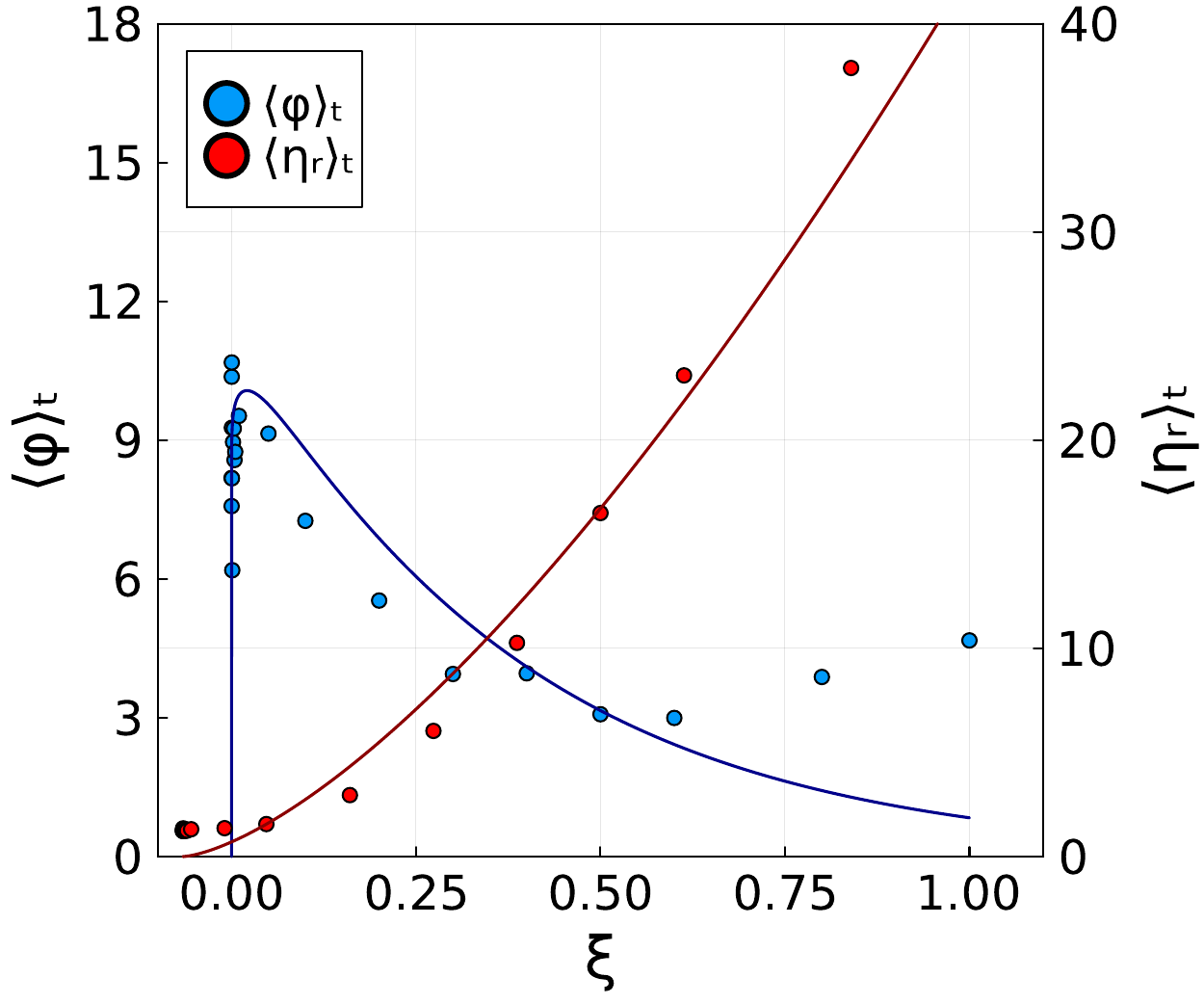}
    \caption{Average fitness and viscosity of agents as a function of noise demonstrate noise-induced order \cite{Matsumoto1983} at low noise levels ($\xi \approx 10^{-4} \sim 10^{-3}$). The average fitness (blue dots) is fitted using an innovation model \cite{Giardini2024}, while the average viscosity (dashed line) follows the short-range component of Sutherland's equation. Both metrics are averaged over time and across all agents.}
    \label{fig:sec4_noise_induced_order}
\end{figure}

Figure \ref{fig:sec4_noise_induced_order} shows the stationary average fitness as a function of the noise level. Low noise levels ($10^{-4} < \xi \leq 4\times 10^{-4}$ rads) are characterized by a sudden increase in fitness. We attribute this phenomenon to noise-induced ordering, a type of stochastic resonance \cite{StochRes} where frequency modes in the noise signal resonate with the main signal, increasing its strength. While we observe patterns such as short segments at noise levels smaller than $10^{-4}$ rads, more complex patterns, such as lanes, emerge at slightly higher noise levels. 

Additionally, the fitness curve is well-fitted by an innovation model that we previously derived using a birth-death Markov process \cite{Giardini2024}. Here, however, noise dictates the evolution of the innovation, forcing agents to quickly adapt to the environment. Strong noise levels, on the other hand, disorient agents, impelling them to abandon the best solution.

Our system is characterized by the absence of long-range interactions, meaning that each agent interacts solely with its nearest neighbors. In this context, the viscosity shown in Fig. \ref{fig:sec4_noise_induced_order} follows the short-range interaction component of Sutherland’s equation \cite{Sutherland} once the noise level, denoted by $T$, exceeds $\xi=0.1$:

\begin{equation}
    \langle\eta_r\rangle\propto\left(\frac{T}{0.56}\right)^{1.47}.
\end{equation}
This behavior is analogous to the viscosity observed in molecular systems with a purely repulsive intermolecular potential $U(r)\propto r^{-2}$, where interactions diminish rapidly with distance \cite{VHS}.

\begin{figure}
    \centering
    \includegraphics[width=0.8\linewidth]{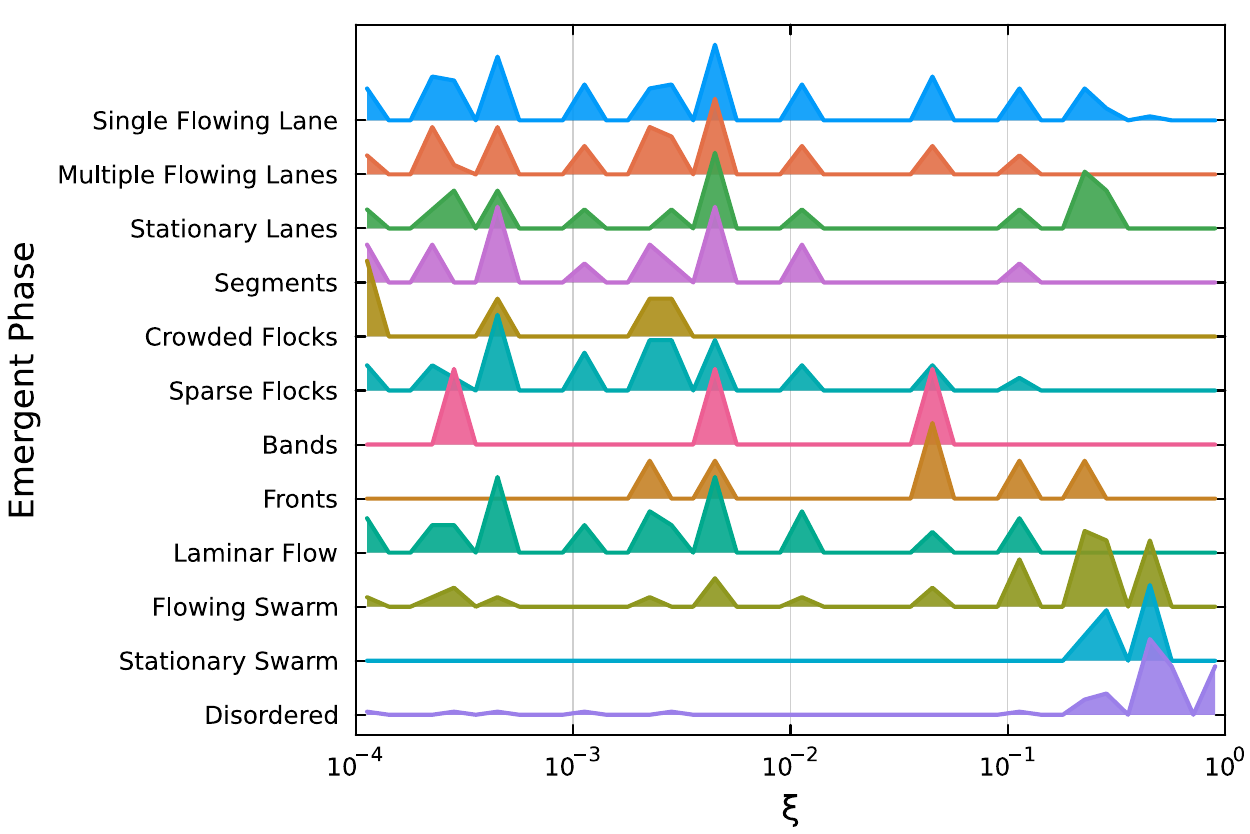}
    \caption{Semi-log (x-axis) ridgeline plot of the probability densities of observing an emerging migration pattern given a noise value $\xi$ shows which migratory behaviors are more robust against noise. Probabilities were normalized by the largest value of all patterns, keeping the same scale for all distributions.}
    \label{fig:Ridgeline}
\end{figure}

Figure \ref{fig:Ridgeline} shows the probability of finding each migration pattern as a function of noise. While certain patterns coexist at various noise levels, clear transitions between different behavioral phases are observed. At low noise levels ($\xi<10^{-4}$ rads), the system is dominated by structured and well-defined migration patterns, such as lanes and laminar flow. As the noise increases to intermediate levels ($10^{-3}\leq\xi<10^{-2}$ rads), more complex and partially ordered structures emerge, including segments and flocks. At high noise levels ($\xi>10^{-2}$ rads), the system transitions to more disordered patterns, such as swarms. Additionally, certain structures, like bands, appear only within specific noise ranges, indicating a delicate balance between order and disorder in the system.

    \subsection{Explaining Agent Behaviors}
    Next, we analyze the internal workings of neural networks within agents, aiming to connect the emergent patterns in collective behavior with the properties of the networks. The goal is to understand how these networks process information. The analysis is performed using a quantitative method (through multilinear regression) and a qualitative visualization of the neural networks' structure. 

\begin{figure}
    \center
    \includegraphics[width=0.9\linewidth]{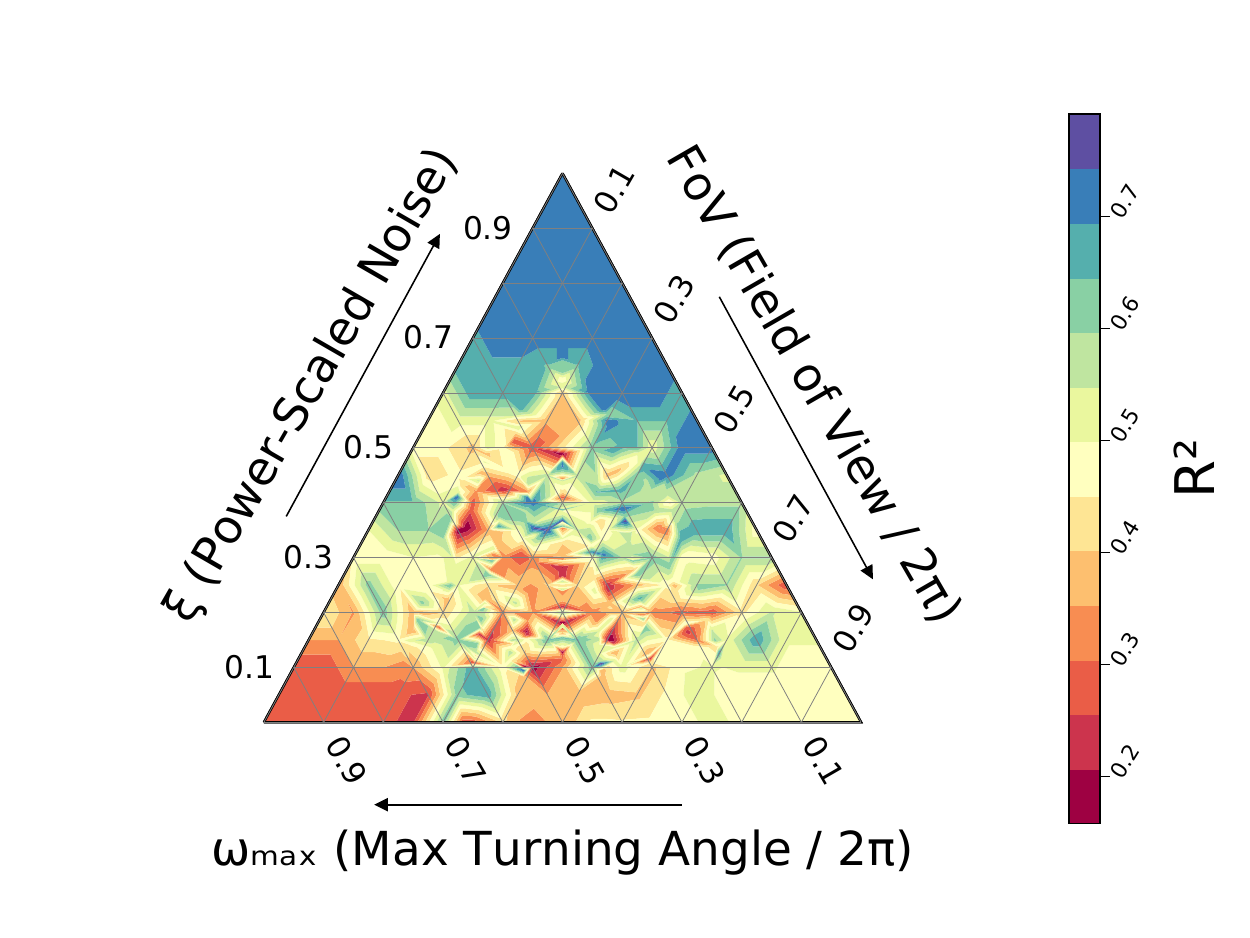}
    \caption{Coefficient of determination $R^2$ obtained for fitting agents' neural networks with a multilinear model depicts different values of non-linearity for distinct emergent migration patterns.}
    \label{fig:sec3_r2}
\end{figure}

For the quantitative study, we use a multilinear regression to assess the relationship between the inputs and outputs of an agent's neural network. Multilinear regression attempts to model the output as a weighted sum of the inputs, essentially assuming a linear combination of input variables. We quantify the goodness of fit using the coefficient of determination ($R^2$), which is the quotient between the explained variance and the total data variance. Values close to $1$ indicate that the network is operating in a relatively linear fashion, while values close to $0$ imply that the network is mostly non-linear. Our data shown in Fig. \ref{fig:sec3_r2} indicates that linear relationships are more likely to be found when agents exhibit simpler and predictable behaviors, such as disordered, lane patterns. Conversely, more elaborate patterns like swarming and flocks correspond to more non-linear networks. 

To complement the quantitative analysis, a qualitative analysis of the neural networks was conducted by visualizing their structure. The neural networks of the fittest agents in each scenario were drawn with a focus on the nodes and edges.

The nodes of the neural network represent neurons, and the color of each node corresponds to its bias value. A color scale was used, where different colors indicated the magnitude and sign of the bias.

The edges between nodes represent the weights connecting neurons. Both the thickness and color of the edges reflect the strength and direction of these weights. Thicker edges represent larger weight values, while the color may indicate whether the weight is positive or negative.

This visualization technique allowed for a qualitative interpretation of how the neural networks process information. By examining the bias and weight patterns, it was possible to infer the relative importance of inputs from neighboring agents and how these inputs were aggregated or prioritized by the network. 

\begin{figure}
    \centering
    \includegraphics[width=1.0\linewidth]{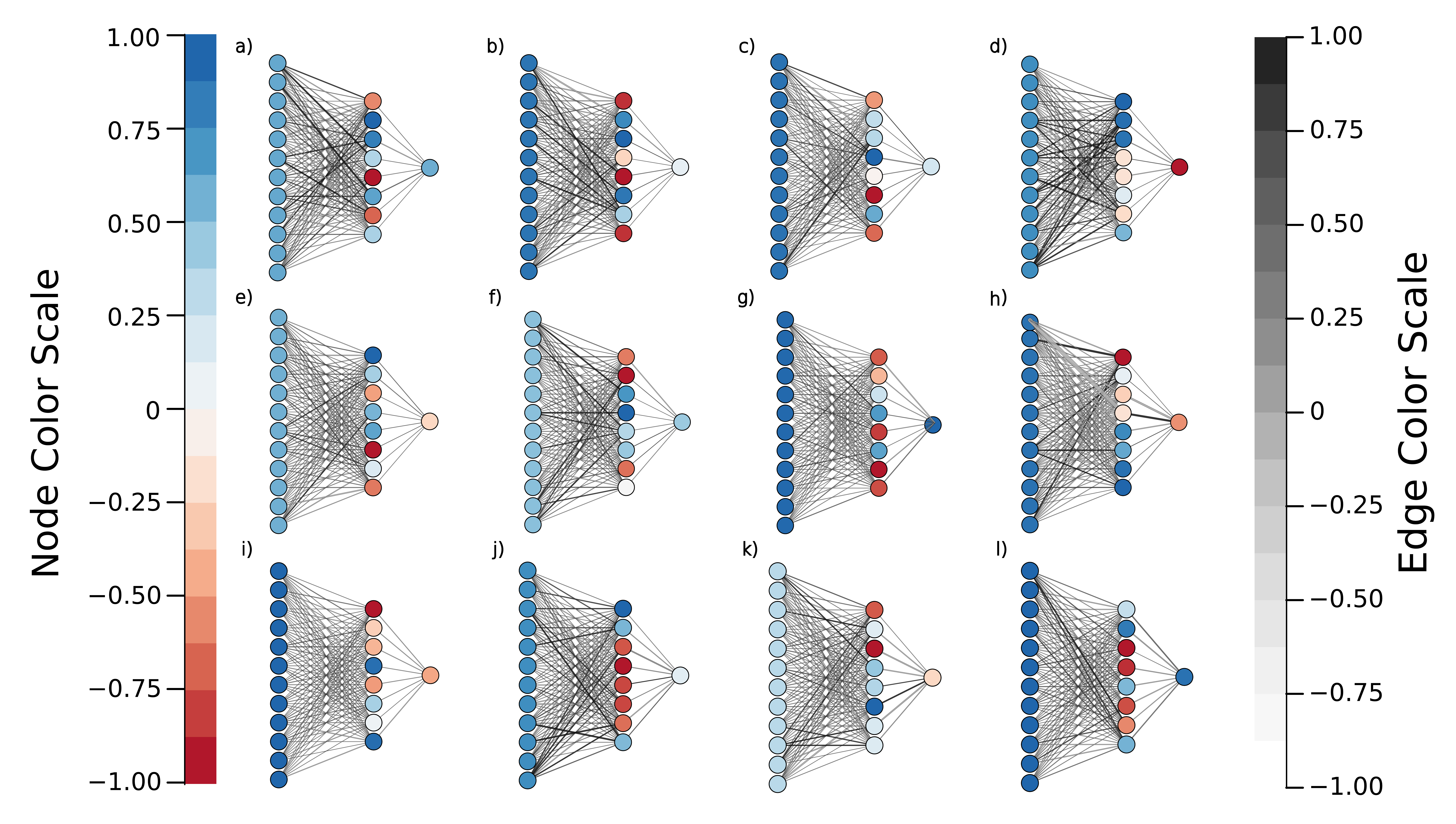}
    \caption{A qualitative depiction of the trained neural networks for each stable migration pattern shows that short- and long-range interactions emerge depending on the observed formation. Nodes are depicted as colored circles with a gradient from red (negative values) to blue (positive values), and edges are shown as lines with a gradient from white (negative values) to black (positive values). The emergent patterns are (a) single flowing lanes, (b) multiple flowing lanes, (c) stationary lane, (d) segments, (e) crowded flocks, (f) sparse flocks, (g) bands, (h) fronts, (i) laminar flow, (j) flowing swarm, (k) stationary swarm, and (l) disordered.}
    \label{fig:sec4_neural_network_graphs}
\end{figure}

Figure \ref{fig:sec4_neural_network_graphs} shows this illustration for different stationary regimes. 
For migration patterns like lanes across the lattice, agents only have to consider neighbors at a very narrow angle in front of them. This corresponds to an increased neural network weight assigned to the nearest neighbors and an averaging behavior for the other inputs. This agrees with the linearity indicated by intermediate $R^2$ values.
For simpler behaviors like bands (as shown by an intermediate spectral gap and lower values of KL divergence), the neural networks average inputs and produce more linear behaviors by assigning similar weights to the nearest neighbors. This is in line with the linearity indicated by higher $R^2$ values.

However, the relationship between pattern complexity shown in Table \ref{tab:sec3_Phase_Complexity} and neural network linearity does not occur for more disordered states such as sparse flocks and swarms. The spectral gap in these cases indicates lower complexity due to the more sparse but well-connected connectivity graph. At the same time, the trained neural networks exhibit high non-linearity and larger weights assigned to more distant neighbors.

\section{Discussion}
The maximum turning angle defines the agility or maneuverability of an agent. When the maximum turning angle is large, agents can make sharper turns and react more dynamically to their surroundings. Conversely, a small maximum turning angle limits the agent's ability to make sharp turns, encouraging more gradual changes in direction.

During the evolutionary training process, agents with a large maximum turning angle are likely to evolve neural networks that prioritize short-term, localized responses to neighbor interactions, since they can immediately adjust their behavior. In contrast, agents with a small turning angle might rely on longer-range coordination to avoid getting stuck in inefficient local configurations. This implies that their neural networks may be selected to process more global information.

Except for laminar flow regimes, where the neural networks are highly non-linear and produce simple patterns, higher values of maximum turning angle are selected for simpler, more reactive neural networks that rely on averaging out immediate neighbors' data. Conversely, intermediate values of maximum turning angle select for more complex and varied migration patterns caused by broader spatial interactions.

Agent density influences interaction range and frequency. In high-density scenarios, agents frequently interact with many nearby neighbors, creating a constrained, crowded environment. Conversely, in low-density environments, agents have more space and interact less frequently.

High-density environments expose agents' neural networks to a high volume of local, frequent interactions. This may favor the selection of networks that can efficiently handle localized, repetitive inputs. These networks might rely more on averaging behavior and shorter-range interactions, which could explain the observed linear behavior in denser environments.

In low-density environments, agents must coordinate over larger distances with fewer direct interactions, leading to the selection of neural networks that process sparser, more complex inputs and prioritize long-range interactions. These networks are more likely to exhibit non-linearity.

The field of view defines how much of the surrounding space an agent can perceive. A wide field of view means the agent can see and process input from a larger number of neighbors, while a narrow field of view, restricts the agent's perception to a smaller region.

High noise levels tend to push the system toward disordered states, where agents interact less cohesively, leading to more random motion. In these regimes, the $R^2$ values are higher, meaning that the neural networks are more linear and predictable. This reflects the simpler interactions where agents primarily process immediate local information and average inputs from neighbors.

Conversely, at lower noise levels, the system tends to favor more organized behaviors like swarming or flocking, which require greater coordination between agents. In these cases, the neural networks exhibit lower $R^2$ values, indicating non-linearity. Here, the networks are engaging in more complex, non-linear decision-making, likely because the agents are processing more intricate patterns of interaction that extend beyond their immediate neighbors.

During evolutionary training, a narrow field of view can produce distinct stable states depending on the noise level. For small noise values, the narrow field of view promotes neural networks that average neighbor inputs to create laminar flow states. At intermediate noise values, neural networks show a preference for distant neighbors and exhibit more complex long-range interactions, likely due to stochastic resonance and limited viewing angles. At high noise levels, the reduced field of view causes neural networks to prioritize nearby neighbors and simpler linear interactions, as frequent orientation changes disrupt long-range connections.

In contrast, a wide field of view forces neural networks to integrate information from a broader range of neighbors, balancing short- and long-range interactions to create intermediate complexity. These broader viewing angles enable agents to engage with more neighbors, maintaining an equilibrium between local and distant interactions.

\section{Conclusions}
In this study, we investigated the emergence of collective behavior in minimally intelligent agents using a minimalistic approach based on evolutionary algorithms and shallow neural networks. By conducting a detailed analysis of the agent's decision-making process, both quantitatively through multilinear regression and qualitatively by visualizing the neural networks, we uncovered key insights into how different environmental parameters shape collective dynamics.

We observed a wide range of collective patterns, including swarms, flocks, lanes, and fronts, emerging from simple evolutionary pressures and local interactions. Our findings indicate that the degree of non-linearity in the neural networks is closely linked to the richness and complexity of collective behavior. Simpler, more predictable behaviors such as stationary swarms and sparse flocks can be largely explained by linear networks that average out inputs, suggesting that these patterns arise from relatively straightforward interactions. In contrast, more complex behaviors like flowing swarms and large-scale flocking are associated with more non-linear behaviors where the networks need to integrate and process diverse, non-local information from multiple neighbors.

We also explored the impact of system parameters, such as maximum turning angle, agent density, field of view, and noise, on the evolution of the agents' neural networks. Our results show that conditions promoting long-range interactions, moderate noise, and intermediate fields of view tend to select more complex and non-linear neural networks, leading to emergent collective behaviors that are more intricate and difficult to predict. On the other hand, both high-density environments and extreme fields of view (whether too narrow or too wide) encourage simpler, more linear decision-making in neural networks.

These insights suggest that to evolve agents capable of rich, emergent collective behaviors, evolutionary conditions must strike a balance between promoting long-range interactions and maintaining a degree of unpredictability through added noise. Moderate maximum turning angles and lower agent densities are also crucial in fostering the development of non-linear neural networks that underpin complex coordination and pattern formation.

In conclusion, this study contributes to our understanding of the fundamental principles underlying collective behavior in multi-agent systems. By demonstrating how complex patterns can emerge from simple rules and interactions, our findings have implications in fields ranging from biology to robotics.

\section{Acknowledgments}

\section{Disclaimers}

\bibliographystyle{unsrt}
\bibliography{main}

\end{document}